# Blockchain Services for Digital Government: An Exploration of NFT Applications in the Metaverse

Zachary Roch[1, 3] and Ramya Akula[2, 3]

[1]MS in Financial Technology

[2]Computer Science Faculty

[3]University of Central Florida, USA

**Abstract**

The full implementation of the metaverse requires the integration of the physical and digital worlds. Applications built on Distributed Ledger Technology (DLT) hold the power to move society closer towards the ideal metaverse through innovations like Non-Fungible Tokens (NFTs). Due to a combination of the infancy of this technology and the significant implications it holds in the public and private sectors, adoption across both sectors is currently limited. To foster the creation of sustainable smart cities built on this technology, education on how this technology may function in an integrated metaverse is paramount. This is due to the necessary compatibility across industries needed between public and private data. As certain industries are more regulated than others, such as finance or healthcare, a robust system is needed to allow for varying degrees of freedom. This chapter illustrates numerous facets of this conceptual framework.

# 1. Introduction

In 2008, a digital transformation began when Satoshi Nakamoto released a paper on a distributed, peer-to-peer fund transfer system known as Bitcoin (Nakamoto, 2008). Since then, words like blockchain and non-fungible tokens (NFTs) have exploded in popularity. Though these ideas caught the eye of the media and captured the public's collective imagination, they are in their infancy while their true potential is yet to be realized. With the ability to use NFTs to tokenize real-world items, avenues for innovation appear everywhere from the event ticketing space to the real estate market. Combining this with the idea of a start-up, NFTs could act as a type of crowdfunding method. Though this scenario could cause all kinds of legal issues as it would be a way for an unregistered company to raise capital, something that is highly regulated within the traditional financial market.

Due to the explosion in the popularity of NFTs, this naturally increased the trading volumes of these digital assets. Extending this line of thought to the convergence of the physical and digital worlds into the metaverse, the ability of NFTs to act as physical items (i.e. government identification card or a delivery of goods), shows that a way to manage and track these transactions and associated data in a secure and verifiable way is needed. This can easily be accomplished through a novel tracking system based on Hyperledger Composer and Hyperledger Fabric (Bal & Ner, 2019), though the use cases for this system are not limited to this. The widespread implementation of NFTs in everyday life is still yet to be seen, especially as they have introduced a slew of new issues and considerations. The complex nature of blockchain technology hinders the implementation and acceptance of it by the masses. For example, people study law for years in order to be an effective contract lawyer. With the introduction of smart contracts (executable code on the blockchain network), functionality of traditional contracts can be replicated on these decentralized networks. In this hypothetical example, for an individual to effectively write a complex, legally-binding, smart contract, they not only need to have the legal knowledge to create the appropriate logic behind the smart contract, but they also need the knowledge of how blockchain networks work and the technical skills to actually deploy the legally binding smart contract.

Regardless of exactly how NFTs will be integrated in the daily life of society, one thing is for certain, as people increasingly work and live together in the metaverse, there needs to be interoperability between the many virtual spaces. One possible solution to this is to use blockchain technology to store each space so that it may be reused and shared (Ryskeldiev et al., 2018). Through the enablement of the above, many people will have the ability to live and work wherever they desire. To make this dream a reality, education of the public is needed. This chapter is designed to explore the concept of NFT services within the metaverse and provide the reader with a detailed, high-level understanding of how they may be utilized in the digital governance of commercial and public entities.

The remainder of this chapter will be broken up as follows: section 2 will create a baseline-knowledge for the reader that the rest of the chapter will be built on top of. Section 3 will give a practical review of various applications of NFTs in the physical and digital worlds and compare between traditional systems and blockchain-based NFT systems. Section 4 will provide a theoretical, in-depth case study that ties many of the themes from section 3 together to extrapolate potential use-cases in a technologically integrated society. Section 5 will explore the limitations and potential risks of NFTs, especially considering quantum computing technology. Finally, section 6 will summarize the findings presented in this chapter.

## 2. Background

### a. Metaverse

In the 1992 science fiction novel, Snow Crash, Neal Stephenson introduced the world to the concept of the metaverse as the characters were transported to a 3-dimensional (3D) virtual world. Building off of this initial idea, the Acceleration Studies Foundation (ASF) released a metaverse roadmap in 2006. In this roadmap, ASF presents a diagram containing two axes, external to intimate and augmentation to simulation, which represent the different types of the metaverse (Smart et al., 2007). This diagram is shown in Figure 1 to show the relation between the four quadrants. Essentially, the metaverse is the convergence, integration, and interoperability of the virtual world and the real world.

One of the key factors that play into the success of the metaverse is the cross-compatibility of systems. In the current state of innovation, there are many "metaverses" or metaverse-like systems, but these are segmented with little or no compatibility between them. Over the years, both individuals and companies have embraced the concept of this shared virtual space, which can be seen in the online video game Second Life (Kaplan & Haenlein, 2009) and the subsequent, continued success of the game Roblox as companies, such as Ikea, legitimize these metaverse-like worlds (IKEA Is Opening a New Store on Roblox, n.d.). While the technological infrastructure is still lacking to deliver the metaverse in its truest form, this has not stopped individuals and companies from working towards this goal, as evidenced by Facebook, Inc. rebranding to Meta Platforms,

Inc. in 2021 (Meta, 2021). To show that the metaverse is beneficial to society across a sample of domains, Kye et al. (2021) explore several applications of the metaverse in educational settings while (Koo, 2021) show that medical education and training can effectively take place from opposite sides of the world. With all of this cross-border collaboration and the rise of digital twins (virtual representations of physical objects, entities/people, or systems recreated within an accurate model of the real-world environment), events need to happen in real-time and in a secure and verifiable manner. Real-time events will need stable, lightning-fast communication; this can be enabled through the continual development of telecommunication systems. As for security and verifiability, this is where the realm of blockchain technology and NFTs excel.

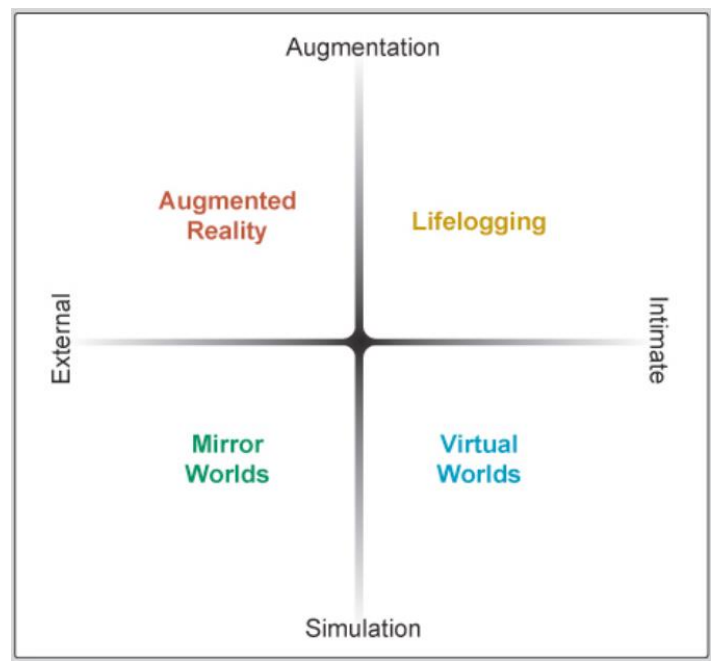

Figure 1: Different Types of the Metaverse (Smart et al., 2007)

b. **Blockchain**

The first blockchain created and widely publicized is that of Bitcoin in 2008 (Nakamoto, 2008). While Bitcoin and other cryptocurrencies like Ethereum (Buterin, 2013) garner significant media attention, the real innovation is the underlying technology behind it; blockchain technology functions as a distributed ledger of transactions where the currency being transferred is a virtual currency native to that blockchain, such as Bitcoin, Ethereum, Solana, Cardano, and many others. The beauty of this type of distributed ledger technology (DLT) is that it provides trust in a trustless environment without the need for a third-party intermediary. It should be noted that blockchain technology is a type of DLT, but not every distributed ledger is a blockchain; however, for the rest of this chapter, the two terms will be used synonymously. For DLT to be effective, it requires multiple nodes (or miners). These nodes will take a bunch of transactions and group them together into a block. Once this is complete, the nodes will compete to solve a complex cryptographic puzzle. The first node to successfully solve this puzzle will then broadcast it out to the rest of the nodes in the network. These other nodes will verify that the broadcasting node's solution is correct before adding the new block to the end of the chain. As each block obtains a consensus amongst the other nodes and is appended to the end of the chain, a predetermined reward is given to the original node that broadcasted its solution (Nakamoto, 2008). During this process, the new block maintains a pointer to the prior block as a way to secure preceding and subsequent transactions, thus leading to its name: blockchain. This means that someone could start at the first block or "Genesis block" and work their way through the entire history until they reach the current state of the blockchain. By designing the system in this way, it provides immutability of a user's transactions as malicious nodes would have an increasingly challenging task of resolving every block's puzzle after the block containing the transaction they want to modify. They will then have to catch up and surpass the current real block while maintaining the majority of the computational power in the network, which is a highly improbable event. Due to the cryptocurrency space being new, there are a high number of scams and fraud being performed by bad actors. However, as regulation, enforcement, and public education of this space continue to grow, the transparency offered by blockchain technology has the potential to hold cybercriminals accountable. Once someone's true identity is tied to a crypto wallet, it is light work to comb through the blockchain to see every single transaction to and from that account. With a little research, it would be incredibly easy for governments and law enforcement to track down ill-intentioned players in this domain.

 Since the introduction of Bitcoin, many other blockchains and cryptocurrencies have entered the scene. Expanding upon Bitcoin's architecture, The Ethereum Foundation launched Ethereum with its native currency, Ether. Embedded within Ethereum is a Turing-complete language which allows additional functionality and customization to transactions. This extra customizability is more commonly called a smart contract and was first introduced by Nick Szabo in 1996. A smart contract essentially adds stipulations to the transaction, such as only sending a predetermined number of tokens once a condition is met. These smart contracts are lines of code that get executed by the Ethereum Virtual Machine, but in order for a contract to be executed, it needs to be funded by an appropriate amount of Ether. This Ether or "gas" is used up during the execution of the smart contract as a type of fee that is paid to the miner node. Through increasingly complex smart contracts, additional versatility is afforded to the network which enables all kinds of new possibilities, such as NFTs.

c. **Non-Fungible Tokens (NFTs)**

First popularized by the online game CryptoKitties back in 2017 (Mani et al., 2021), NFTs are digitally unique, verifiable, and cannot be subdivided into smaller parts like cryptocurrency which can be divided. This one-of-one characteristic is what really sets NFTs apart from other fungible tokens (i.e. Bitcoin or Ether) and provides a great utility value for them which will be covered later. Though NFTs can exist on different blockchain networks, such as Solana and Cardano, where each network sets forth their own respective standards for the tokens, in order to focus on the concepts in this chapter, the Ethereum standards mentioned here are an illustrative example only. The Ethereum network allows for other cryptocurrencies to be built on top of it. These tokens must follow the ERC-20 standard which is essentially the set of rules that provide for full fungibility of the tokens. NFTs, however, follow the ERC-721 standard which provides for its nonfungibility. Another notable standard is ERC-1155, which provides for the semi-fungibility of tokens.

Another key characteristic of NFTs is that they can contain much more than simple transaction information, data such as pictures, video, or even sound. This subsequently opens the door to blockchain-based video games. From here, NFTs saw a meteoric rise in popularity due to the success of the game CryptoKitties (CryptoKitties, 2019), which garnered over 70% of the transaction capacity on the Ethereum network (Mani et al., 2021). This outcome, though spurred on by innovation, is particularly alarming because the necessary amount of gas paid by entities on the network increases as the available transaction capacity decreases. This creates the issue where a smart contract may fail to fully execute should it not have adequate funding of Ether. This excess friction directly results in additional money spent by end-users, a situation that would slow the adoption of blockchain technology. At this time, Ethereum was running on a "Proof-of Work" (PoW) system, but it has since switched to a "Proof-of-Stake" (PoS) system (Kapengut & Mizrach, 2023). The main distinguishing factor between these two consensus mechanisms is that blocks added under a PoW systems require verification by a number of nodes on the network via a complex cryptographic puzzle while blocks added under a PoS system are essentially verified by a node that stakes a portion of their available Ethereum to vouch for the validity of the new block. This directly resulted in energy consumption decreasing, daily transaction numbers increasing, and a lower concentration on the network (Kapengut & Mizrach, 2023).

3. **NFT Applications**

   The primary goal of this section is to review NFT applications across several digital and physical markets which can replace or enhance various aspects of the digital or physical world respectively. This is accomplished through a comparison of traditional and blockchain-based solutions. It should be noted that due to the integrated nature of the metaverse, there will be some overlap between the two sections.

   a. **Digital World Applications**

      In December 2021, the digital artist known as Pak released their NFT, Merge, for $91.8 million. This marked a new record for the most expensive artwork sold by a living artist, though it is debatable on whether this piece can be considered a single art piece (Crypto.com, n.d.). As one of the arguably most well-known applications of NFTs so far, the digital art and collectibles market is positioned well to lead this discussion. Art, by definition, is unique with no two pieces being the same, this subsequently makes it a great fit for NFTs. By tokenizing digital art pieces and embedding them with a unique hash on the blockchain, it creates an immutable record of ownership (Rehman et al., 2021). Since a record of the transfer of ownership of the art or collectible is maintained on the distributed ledger, it allows anyone to verify whether a given party is the true owner of the piece. Another equal or potentially greater benefit to this market is granted due to NFTs being the product of smart contracts. This allows artists to embed a commission into the NFTs smart contract so that whenever the piece is sold on the secondary market, the artist earns a residual commission. These features combined revolutionize this market compared to the traditional system; digital

artists can passively (depending on the exact specifications of the smart contract) collect residual checks for past work while not needing to worry about their art being stolen and reposted by another as an original work.

Another group that quickly adopted NFTs and were contributing members to their explosion in popularity would be the gaming community. Merging blockchain and gaming is almost a natural fit as this population tends to be more technologically adept, a trait necessary in the current phase of blockchain technology. In video games, NFTs can take the place of items, characters, trophies, and more. With the introduction of dynamic NFTs (dNFTs), NFTs that can change or update their metadata overtime based on some type of input (Guidi & Michienzi, 2023), as events take place within a game, the dNFT will update automatically to reflect the change. On top of this, dNFTs open the door to cross-game (and subsequently cross-metaverse) interoperability, as players may be able to mint (create) a dNFT during normal gameplay and then transfer their item to a new game where it can retain or modify the item's attributes. In the case of certain metaverses, NFTs are used to track the exchange of virtual land (Dowling, 2022). A perk of the immutable nature of blockchain technology is that even if a game studio closes, the player will retain any dNFTs that were tied to the game. They may either keep them to display, use elsewhere, or they can even monetize their items by selling them through various online brokerages such as OpenSea (OpenSea, n.d.) or Rarible (Rarible, n.d.).

Though this integration is not seen fully yet, games such as the anticipated Pudgy World (Penguins, n.d.), push the boundaries of the current metaverse as they usher in the integration of physical toys and digital assets in an immutable way. Users can buy a toy at their local Walmart which will come with a QR code that allows them access to unique traits for their in-game 'Forever Pudgy' (Malwa, 2023), similar to how activating a Webkinz toy online works (Welcome to Webkinz World!, 2019). When compared to traditional video games, the interoperability and permanence of blockchain-based games provides a significant enhancement to the industry. In their current state, blockchain-based games as a whole currently lack ease of access, playability, and design when compared to traditional AAA games (a designation for large, well-funded, and highly anticipated games). At its core, gaming is meant to bring enjoyment and while DLT holds great potential for the future of the gaming industry, the overall quality and the number of games pale in comparison to the traditional video game market.

In a similar vein, the content creation and distribution sector is susceptible to this innovation. Currently, individuals and businesses use various social media platforms to publish their content and interact with fans. Some parties will create free content, on platforms such as YouTube, as well as content that is locked behind a paywall, on platforms such as Patreon, to entice people to make a purchase. While this is great in theory, it does little to stop people from sharing the content with others once it is in their possession. Additionally, the content tends to be unidirectional where the creator makes the content, and the subscribers consume it. Since NFTs can act as containers for digital content, this allows individuals and businesses to restructure how they create and deliver their content as well as how they interact with their fans. For example, exclusive or higher quality content can be locked away behind the NFT so that only the holder of the NFT can view or use the content. By placing the content within NFTs, it is easier for the creator to track who is using it and be fairly compensated for their work. Two great examples of NFTs in use within this space are Fox Entertainment's company, Blockchain Creative Labs, and Dapper Labs creation of NBA Top Shot (Wilson et al., 2022).

Looking at the intersection of content creation, 3D modeling, and fundraising, a unique use case for NFTs can be seen to assist in the modeling of museum pieces. By nature, 3D modeling can be expensive both financially and computationally. This poses a challenge to museums who would need to create countless models for all of their exhibits. Through an initiative to model every exhibit, a museum could immortalize the exhibit as an NFT. Not only will this help preserve the attributes of the exhibit piece in case it is accidentally destroyed or lost, but it will also provide a learning opportunity for people who are not able to see the exhibit in person. Through linking the exhibit piece to an NFT, the museum can retain the copyright ownership while selling the NFT to offset the 3D modeling costs (Bolton & Cora, 2021).

b. **Physical World Applications**

Similar to how NFTs can tokenize digital assets, there is an incredible potential advantage to the tokenization of real-world items. Drawing on the similarity between the digital art/collectible market and the physical one, NFTs may play a crucial role in the improvement of the authentication of artwork under the current system, as it "attaches a unique hash with each piece of art that allows it to be differentiated" (Rehman et al., 2021). While many NFT collections on the market are algorithmically generated with pieces in a collection generally sharing various characteristics or theming which is determined upon minting the NFT, tokenizing real-world art would function slightly differently due to its existence prior to NFTs. One potential way this could work would be for the owner to take their piece to a professional who would verify its authenticity. The third-party authenticator can either be the same party that mints the certificate of authenticity as a NFT, or they can partner with a pre-existing platform for this step of tokenization. Linking the art to NFTs provides the additional benefits of preserving the work should it be destroyed as well as the potential to reduce the instance of fraudulent art sales.

    Drawing further inspiration from the tokenization of digital land in the metaverse, there are several ways that NFTs can be applied to the physical real estate industry (Dowling, 2022). An NFT and an individual's house share the same property of being unique; two identical looking houses next to each other are distinctly different as they are on two different plots of land. The tokenization of real estate provides for some interesting characteristics and possibilities as it provides for a public, immutable record of ownership which can increase the efficiency of title searches, reduce transaction friction, and eliminate the issue of lost, destroyed, or stolen titles. With the introduction of composable NFTs, a way to group an arbitrary number of NFTs and fungible tokens (Guidi & Michienzi, 2023), a large investment property can even be tokenized with multiple owners (Despotovic, 2022). Each owner would hold the composable NFT as a type of

membership with the nested tokens representing the share in the property. This type of collective membership via NFTs can also be called a Decentralized Autonomous Organization (DAO). A DAO is essentially a blockchain-based governance system used to disseminate voting power, management, and ownership of the underlying asset/s. This can allow for small-time investors to pick and choose which geographical markets and individual properties they want to invest in without the need for investing into a Real Estate Investment Trust or other more traditional investment vehicle. Compared to the current state of the real estate market, the introduction of NFTs can be seen as a double-edged sword. While there are many benefits summarized above, the tokenization of real estate, especially with the introduction of DAOs and the ability to trade in localized markets by international players, is set to provide a plethora of regulatory/legal issues. An in-depth look into some of these challenges will be covered later in the chapter.

    The tokenization of real-world assets can extend beyond single family residences to include cars or other high-ticket items, such as high-end fashion (Rehman et al., 2021). Buying a house or car is often one of the largest expenses in a person's life, and for the most part, there are many moving parts that take place for the successful transaction. A couple of the home purchasing steps include a title search, securing funding, inspection, negotiating the price, etc. When buying a car, some or all of the above steps may be conducted, all of which can eat up a large chunk of time and monetary resources. Should large-ticket items be tied to NFTs and the whole transaction process be put on the blockchain,

it can reduce the burden on both the buyer and seller. Consider this, in a fully integrated society, any repairs, modifications, service, etc. on the asset in question can be recorded within the dNFT. This provides a track-record of the asset's condition. Just by inspecting the NFT, it can be determined if there are any liens on the asset, when a new roof was put on (for houses), when the last oil change occurred (for cars), or what refurbishment was completed (for damaged high-end fashion or art). This provides more accessibility and transparency to information during the transaction. In the process of seeking funding for the asset, the lender can also view all of the information on the asset as well as the buyer and seller's relevant personal information held within their "Soulbound NFT" (Weyl et al., 2022). Briefly, a soulbound NFT is a non-transferable NFT that contains various traits of the holder. This helps streamline the time and cost to underwrite the asset as it will make the process more automatic, there will be less back-and-forth between the bank and the buyer to seek out documentation and personal financial statements. Through the additional functionality of smart contracts, once the lender underwrites the loan, they can wait to release the funds directly to the seller until both parties agree to the transaction. Before being presented with the option to agree to the sale, all of the stipulations within the smart contracts must be met. For example, a used car purchase is taking place. The buyer specifies that the tires must have a certain amount of wear left on them. In order to satisfy this condition and record its validity, the seller can upload proof to the car's dNFT.

  With the use of soulbound NFTs, the traditional real estate and financial markets can be completely transformed. In an integrated metaverse, the tokenization of real-world assets can be extended to include traditional investment vehicles. This decentralized ledger would have to be interoperable with the notion of traditional fiat currency. When a Central Bank Digital Currency (CBDC) (Lovejoy et al., 2022) is introduced as a potentially more trusted currency than some of the stable coins that exist today, this can help bridge the gap between the current opaqueness of regulation of digital assets and the intense regulation of the traditional financial markets. A detailed case study is presented in the next section that expands on this idea to focus on the implications within the realm of finance and regulatory agencies.

    With one of the largest event promoters, Live Nation, currently under fire for anticompetitive practices (McCabe & Sisario, 2024), DLT is set to disrupt the event ticketing industry. Currently, the process of seeing a popular artist is blurred at best. Generally speaking, a ticket is not purchased directly from the artist, but rather through a third-party site, such as Live Nation's Ticketmaster, who is free to charge additional fees. For highly anticipated shows, tickets occasionally sell out; this provides an added layer of risk to the person who still needs to purchase a ticket. If a ticket for a sold-out event is purchased from someone online, the seller gets to dictate how much the ticket now costs and there is no guarantee of the ticket's legitimacy. The verifiability of NFTs provides a level of trust for purchases in the secondary market. On top of this, tickets issued as NFTs allow the artist to specify any number of conditions to control sales in the secondary market (Regner et al., 2019), such as only being able to sell the ticket back to the artist, setting a cap or floor on the price of the ticket in the secondary market, and

being able to collect a residual fee (similar to the art market) with each subsequent sale. The artist can release their tickets as semi-fungible tokens, so that buyers can exchange like tickets (within the same section) with each other prior to the show. After the show, the token will transform to become non-fungible as it loses its utilitarian value (unless otherwise specified) and retains its collectible value.

 As society becomes more integrated, education is bound to change to adapt with technology (Kye et al., 2021). Currently, primary education is heavily limited to geographical location, though this tends to change for people that pursue college. A person's residence generally determines where their children go to school. This means that someone's child may receive a less desirable education that does not correspond with their intelligence level. By integrating blockchain, NFTs, and digital twins, education can be accessible for anyone anywhere as each virtual classroom can be stored digitally and reused as needed (Ryskeldiev et al., 2018). This can become especially helpful for classes that involve a lab component. By utilizing augmented and/or virtual reality, a hands-on experience can be delivered to more people. Transcripts can even be tied to soulbound dNFTs to create a verifiable record of someone's education and ease the financial burden of sending transcripts to every college that they apply to. While DLT can provide an immense benefit to education, one of the largest downsides of it is the accessibility and technical knowledge, as users will need access to the appropriate equipment, have very fast connectivity speeds, and the technical know-how to operate in a decentralized space.

Similar to how soulbound dNFTs could work with a school transcript, this opens the possibility to their use within the healthcare industry. Since NFTs can be seen as containers that house digital information, they can be used to securely hold patient data (Skalidis et al., 2022). A great use-case is when someone moves to a new town and establishes themselves with a physician. There is a lot of paperwork that must be filled out and the new physician must work with the patient to receive the patient's medical record from their old doctor. After this entire process, the patient doesn't know for sure what their old doctor did with their medical records, whether they are being retained for a period of time, securely disposed of, or if they are even locked away. If the new physician prescribes the patient with medicine, the prescription must be sent to a new pharmacy too. With the ability of NFTs to be tied to oneself, using smart contracts, a patient can grant access to medical staff for an allotted amount of time or until some other condition is met. This puts the patient's data back into their hands so that they may manage it appropriately. Whatever work is performed on the patient at the hospital can be billed to their virtual wallet or their health insurance company's wallet. An itemized list of expenses can be made standard practice to be uploaded within the dNFT. This can help reduce frivolous medical expenses and speed up the claims process as, in a perfect world, all three parties will have all the information they need without contacting each other. NFTs can also be used in healthcare in combination with digital twins in order to recreate and run simulations on patients, equipment, or even the entire hospital (Musamih et al., 2022). They can also be used within the medical and pharmaceutical supply chains to provide additional transparency and reduce the amount of counterfeit equipment or medicine (Musamih et al., 2022), but this logic can be extrapolated to any industry supply chain that lacks transparency or suffers from counterfeit items.

Another application of NFTs that can be applied to nearly every industry is found in the use of Intellectual Property (IP) protection. While someone is engaged in the process of obtaining a formal patent, they may use an NFT to verify that the idea and its details are the true property of the creator (Bamakan et al., 2022). Regardless of when the person receives their patent, tying their invention to an NFT allows them to profit off their invention easier through reducing transaction frictions when selling their patent (Fairfield, 2022) or through ease of leasing their patent to collect residuals checks, like how NFTs are used with digital art.

## 4. NFTs, Digital Government, and Financial Technology

By using short examples in section 3 to explain the concepts behind various NFT applications, an in-depth theoretical case study is presented here that blends the above topics together to illustrate their potential roles in the context of a fully interoperable digital government within the metaverse. Though there are several different definitions of what a digital government is, this discussion will use Robertson and Vatrapu's (2010) definition that digital government, "encompasses the use of information and communication technologies to enable citizens, politicians, government agencies, and other organizations to work with each other and to carry out activities that support civic life". To classify the stakeholder interactions within a digital government, there are government-to-government (GtG), government-to-business (GtB), business-to-government (BtG), citizen-to-government (CtG), government-to-citizen (GtC), and citizen-to-citizen (CtC); the remaining pairings belonging in the realm of e-commerce (Robertson & Vatrapu, 2010). The following sections will give an overview of how NFTs can be incorporated into the normal interaction of the above stakeholder pairings. Finally, a financial technology (FinTech) example will be given that ties everything together.

### a. Identity Verification

Within any governmental system, identity (ID) verification and management is of the utmost importance, as is evidenced by voter fraud. When combined with ID verification, blockchain technology and NFTs provide an excellent way to cut down on instances of manipulation with voter fraud (Casado-Vara & Corchado, 2018), and it can create a positive impact on the security of democratic processes (Olaniyi, 2024). Since NFTs are built on the blockchain, an immutable record, they provide a viable alternative that solves the identity verification issue. Until recently, the public's collective attention has been focused on NFT applications that are transferable. This is due to the initial popularity of the art/digital collectible market with the trading of such assets entering the realm of its own market with associated price indices (Schnoering & Inzirillo, 2022). In contrast to this transferability, soulbound NFTs (Weyl et al., 2022) provide a digital solution for the various hard-copy government identifications such as a birth certificate, a driver's insurance card, or a medical record. A person can then grant permission to necessary individuals to access specific records contained by the NFT with a built-in mechanism, such as Proxy Re-Encryption (Agyekum et al., 2021), for allowing certain government authorities and medical personnel to access their respective documents in extenuating circumstances. This would allow a person to set preferences for who/what can access each document and under what circumstances. Looking at the various stakeholder relationships (GtG, GtB, BtG, CtG, GtC, CtC), NFTs can be implemented in a plethora of ways. Some examples include a secure digital signature when sending classified documents, automatic verification of federal employees when conducting government business, or even as a way to prevent scams on online marketplaces (i.e. Facebook Marketplace). See Figure 2 for a breakdown and visualization of how these stakeholder pairings may use NFTs.

|  | Government-to-Government (GtG) | Government-to-Business (GtB) | Business-to-Government (BtG) |
|---|---|---|---|
| Identity Verification | A secure digital signature when sending classified documents between agencies. | Verify the business and senior management in order to facilitate regulatory correspondance as in the case of a bank or restaurant. | The business can verify itself in order to follow-up on formal enforcement actions. |
| Public Records Management | A record of court proceedings with all relevant supporting documentation and judgement. | A secure and verifiable way to manage all active and past government contracts with various businesses (such as defense contracts). | A way to upload and manage company filings with the Security and Exchange Commission (SEC). |
| Public Service Delivery | Documents (classified or not) can be held within an NFT for the review of different people or agencies. | Notification, status, and relevant documents for the forgiveness of the Paycheck Protection Program loan can be stored within an NFT. | Streamlined way to file for a new patent. |

|  | Citizen-to-Government (CtG) | Government-to-Citizen (GtC) | Citizen-to-Citizen (CtC) |
|---|---|---|---|
| Identity Verification | Provides a way to ensure one's identity for the purpose of applying for a passport. | Can be used by first responders to identify someone and provide medical care. | Can be used when purchasing a good from someone online, through platforms like Facebook Marketplace, in order to verify that the person and item exists. |
| Public Records Management | Provides a way to store all past tax returns in the event of an audit. | Provides a way to securely store the title of who owns each plot of land. | Can be used in place of a title search to verify the ownership and status of a car/house in a private sale. |
| Public Service Delivery | Streamlined way to file for a new patent. | Issuance of a new driver's license can be sent to a digital wallet as an NFT. | The NFT for the title of a house/car can be transferred directly to someone in the event of a death. |

Figure 2: Example Applications of NFTs by the different stakeholder pairings

b. **Public Records Management**

An important aspect of any government system is that of managing the copious amount of paperwork and documents. For example, every plot of land or car has a corresponding title; every misdemeanor conviction also carries with it corresponding documents. Though more likely for private sector data, it is possible that these documents are not currently secure and are open to being tampered with or they could go missing somehow. Tying these documents to NFTs can resolve this issue. In the case of the automotive industry, composable dNFTs allow citizens to keep a digital copy of their title with them at all times. Continuing this thought, NFTs can be used instead of a driver's license, passport, or anything that requires some type of identification which would enable the creation of a smart city (Musamih et al., 2022). Per IBM, "A smart city is an urban area where technology and data collection help improve quality of life as well as the sustainability and efficiency of city operations" (Gomstyn & Jonker, 2023). Since all transactions are recorded on the distributed ledger, it has the added benefit of reducing the chance of fraud (Mozumder et al., 2023). See Figure 2 for a more in-depth look and visualization of how the stakeholder pairings may implement NFTs in public records management.

c. **Public Service Delivery**

The third pillar for any entity, public or private, involved in digital governance in some capacity is public service delivery, which is ripe for innovation coming from DLT. The distributed and digital nature of blockchains and NFTs streamline the public service delivery process thus improving the overall experience of the citizen. They reduce the cost and time of the administrative process, as agencies can issue digital licenses, certificates, or patents (Bamakan et al., 2022). The documents contained within these NFTs can then be linked securely to the citizen's own digital identity via the soulbound NFT. Additionally, NFTs can be used to streamline the personal accounting process. If government identification can be held as a NFT, it follows that tax returns, or any sensitive information, can be tied to an NFT and submitted that way. With the introduction of a CBDC (Lovejoy et al., 2022), filing taxes has the potential to be automatic with the tax due/credit being directly debited/credited to digital wallet via an airdrop (Allen et al., 2023). An airdrop feature can have additional implementations as it could be used to deliver government benefits, such as social security payments or food stamps. A dNFT can be issued to someone that conveys the person's need for the benefit, as well as what benefits are available. The person would then have to go to the store to redeem their benefits for real-world items, all of which is recorded in the ledger! Within the metaverse, the person may choose to not even leave their home to shop, but rather redeem their benefits in a virtual world to receive physical goods. Once the person no longer qualifies for the financial need, the NFT can be burned via a smart contract. The burning of a NFT is the removal of it from the overall supply, a burned NFT can effectively be considered destroyed. For a more in-depth look and visualization of NFT applications used by the stakeholder pairs in the public service delivery process, please see Figure 2.

### d. Theoretical Case Study

The impact of NFTs and blockchain technology can have a great effect on digital government and the citizen experience. They have the opportunity to enhance identity verification, records management, and distribution of public services by improving transparency, efficiency, security, and the overall trust of the public. Within the realm of finance, these properties are of the utmost importance. This is due to the extensive regulations set forth in the United States for traditional financial products. With decentralized finance regulations still in their infancy, the following theoretical example sets the issues that arise from this technology aside for the time being. As there are currently a limited number of practical examples that help illustrate all of these concepts as well as a lack of seamless interoperability between the digital and physical worlds, this example should be viewed as a type of framework where the interactions that take place have the opportunity to be the future outcome of the work accomplished in this field today. Many of the following interactions within this example can be seen as building blocks that use the concepts presented in this chapter, where each building block has the potential for some innovative business venture. This example will follow an independent singer-songwriter named Satoshi (after the pseudonymous creator of Bitcoin).

Satoshi is scheduled to perform a show later this evening. Leading up to the performance, he accessed digital twins of two different venues stored within the blockchain to simulate his concert to find the optimal venue that accentuates his music style. He then booked out the venue, hired security, and hired a production team. When he booked the venue, he received a dNFT which acts as his key and identification to get into the venue. His identification is stored within a soulbound NFT so that he can access it anywhere. When he hired security and the production team for the event, the funds used to pay for them were essentially locked up by the specifications of a smart contract. The teams will receive their funds once all stipulations of the contract have been fulfilled and both parties sign off on the transaction. A way to arbitrate disagreements is needed in case either party doesn't fulfill their obligation. This could be a type of decentralized insurance policy or more directly through an oracle, a way to input off-chain data into the blockchain. The tickets Satoshi sells are in the form of semi-fungible NFTs. With each ticket, not only does the buyer get access to the event, but they also receive a physical T-shirt and a live recording of the event once it is complete. When concertgoers arrive and scan their tickets to enter, the fungible token is converted to a non-fungible token which allows them to pick up their shirt at the merchandise booth. After the concert, the attendees can either keep their NFT for the collectible value and the utility value of rewatching the concert (data that is accessed via the blockchain) or they can sell it to recoup part of their initial expense. Satoshi wanted to ensure that his shows are accessible, so if anyone were to sell their ticket before the show, it would return to Satoshi at a predefined price, but once the show starts, any additional sales of the now non-fungible token are not set to return to Satoshi. After the concert, Satoshi stops to pick up some food. The restaurant, which he is a reward member at, will scan his wallet in order to add reward points based on a membership NFT and deduct the appropriate amount of the CBDC needed for the food.

    Through many years of hard work, Satoshi's music career starts to take off. He now has more money than he knows what to do with. The first thing he does is to purchase an insurance policy on himself. To underwrite the policy, he allows access to his medical records by the insurer. This insurer is a little unique as it is made of a DAO. Thankfully he is in good health and the participants of the DAO issue the policy. He then turns his eye to option trading in the financial markets. He has never been in a position to open a brokerage account before, so he must go through the Know Your Customer (KYC) process as well as the application to trade on margin within the account, where the account is structured as a type of composable NFT. He heard that this process used to be cumbersome and repetitive at each firm, however it was explained to him that he only needed to complete this once before he can freely open accounts at various brokerages using the identification held within his soulbound NFT. On a random Saturday, Satoshi is presented with the opportunity to pair up with several other investors to purchase a large investment property. These investors come together to form a DAO to manage the property. To liquidate the funds for the purchase, Satoshi sells a few index funds over the weekend and can instantly transfer it to purchase the property the same day, a feat never before possible in the previous financial market due to the settlement time of equity trades. After holding on to his share of the property for a while, he decides that the real estate market isn't for him, so he

puts it up for sale in a decentralized exchange where another investor purchases it several days later.

Due to his recent success, Satoshi is the target of a lawsuit. On the day he is supposed to appear in court, he walks into his office, turns on his computer, and appears before the judge in a virtual reality environment. His case is recorded and the court transcript is stored in an NFT. During the actual court hearing, the judge pulls up the defendant's profile and personal information held in an NFT. Thankfully, Satoshi is found innocent as a guilty verdict would be put on the record in his soulbound NFT (Weyl et al., 2022). Since the support for multiple virtual spaces can be put on the blockchain (Ryskeldiev et al., 2018) and accessed as needed, his case ends up being used later on to teach a legal studies class.

For particularly interested readers who wish to take this a step further: think about what is accomplished during the day-to-day activities of the personal and professional life. Perhaps a few things will jump out that occur with higher frequency or maybe there is some task that brings a level of frustration. Then start to play around with the concepts presented here to see if any of them may have a viable use-case at the intersection of need and feasibility. To see a visualization of how financial transactions take place in an integrated metaverse, please see Figure 3.

## 5. Challenges/Limitations

While much of this chapter is focused on the potential benefits and innovation that NFTs can bring to digital government in the metaverse, it would be in poor taste to pass by the many obstacles in this field. For something to be widely accepted and commonly used, technology giants, Meta and

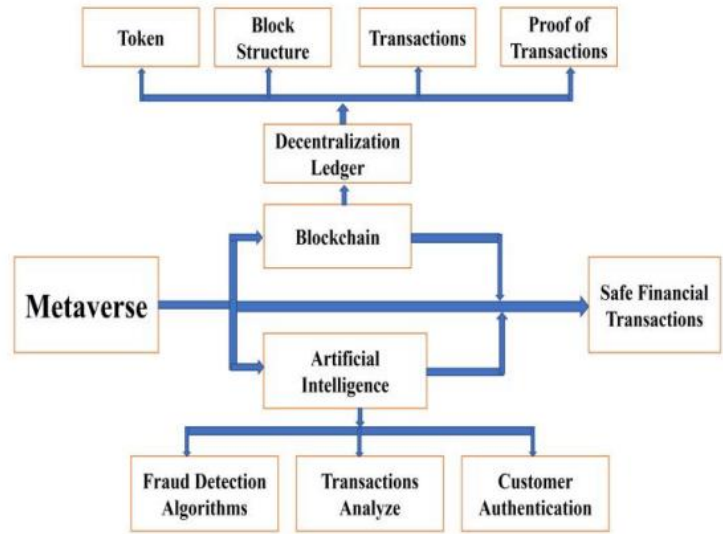

Figure 3: A Tentative Metaverse Approach for Financial Transactions (Mozumder et al., 2023)

Alphabet, who provide Facebook accounts and Google Search respectively, provide a great example of this acceptance. While the common person likely won't know the technical details behind how Facebook's friend suggestion page works or how Google's search engine works, they will likely know how to use each service and what benefit it provides. This means that for blockchain and NFTs to become widely used and accepted, products and services integrated with this technology will need to be easy to use and understand by the common end user. From the consumer's point of view, NFTs and blockchain are very complicated topics. The average person may find it rather difficult to figure out how to set up a wallet or transact on a network, let alone learning about how gas works and how to mint an NFT. This establishes a need for further education of the public and better interoperability and seamlessness in the end-user experience. The complexity of the blockchain hinders this goal for developers as well. While this group of people are generally more knowledgeable on the specifics of technology than the normal person and thus may learn the technicalities of something new in this realm easier, it is still not an easy feat to obtain a thorough, working understanding of how blockchain technology works. After this, there remains the issue of the developer implementing this technology seamlessly and distributing it to the end users. To assist with the transparency of smart contracts and thus the reduction of failed contracts, developers and blockchain explorers can indicate environment variables, provide adequate detail for transaction failures, and others (Hartel & van Staalduinen, 2019).

In a fully interoperable metaverse which employs blockchain services, great internet connection is to be desired. Due to the nature of the blockchain, actions someone takes online occur almost in real time. For example, when a user is getting rid of Ether, the transaction is processed shortly after placing the trade; that same user can then place a subsequent buy order followed by a sell order and the funds will be in their account, ready to use for the next transaction. This can also be called on-chain settlement; however, work is being done to facilitate cross-chain settlement in an interoperable environment (Lee et al., 2021). Due to this near real-time nature, participants would benefit greatly if there was a stable and fast internet connection for them to use. Though 5G network technology is beneficial, this can hopefully be provided in a more complete sense in the medium to long-term by 6G network technology (Jiang et al., 2021).

Another obstacle facing the adoption of blockchain technology and NFTs is its interoperability with legacy systems. This is needed for both small, private blockchains and large community-based projects. There may even be the need to make the private blockchain work with the public one (to some extent). This issue will be seen by nearly every industry from healthcare (Musamih et al., 2022) to banking. Taking both of the aforementioned sectors as an example, there is a lot of information involved in both industries. This poses another issue as high storage requirements and frequent transactions can quickly clog up the blockchain, resulting in higher gas fees and a delay in the confirmation of a transaction. This becomes especially pertinent in the implementation of a smart city, as a multitude of transactions are completed by nearly every person, every single day. Therefore, a blockchain that has a high storage capacity and a high throughput is needed. Regarding the former, an InterPlanetary File System (IPFS) may be implemented (Zheng et al., 2018), though this raises an additional concern surrounding the immutability and longevity of the off-chain documents. Regarding the latter, advances such as Ethereum's PoS system are helping to increase the transaction throughput. Other technical challenges remain, like security via smart contract vulnerabilities. These could result in minor issues, such as the loss of a small gas fee due to a failed contract or the issues could be much larger, such as someone illegally obtaining a person's NFT containing their sensitive medical information.

Another significant threat to the longevity of blockchain technology comes in the form of quantum computing. The success of many current blockchains is built on the back of the Elliptic Curve Digital Signature Algorithm (ECDSA) and the SHA-256 hashing algorithm for their security. In light of the advent of quantum computing, these security features can be broken by a sufficiently powerful enough quantum computer running either Shor's (Shor, 1994) or Grover's (Grover, 1996) algorithms. With the current state of many popular blockchain networks being susceptible to a quantum computing attack, research of and the eventual transition to a Post-Quantum Blockchain (PQB) is of the utmost importance. Thankfully there are already post-quantum digital signatures for transaction verification, consensus mechanism, and communication schemes where many of these proposed solutions are built on quantum-resistant, lattice-based cryptosystems, where the algorithms are based on structures consisting of points in n-dimensional spaces that repeat (Gharavi et al., 2024). An example of one of these solutions is Ethereum 3.0. Under Ethereum Improvement Proposal (EIP) - 4844, security and scalability can be improved through the use of Proto

Danksharding, and quantum-resistant elements known as Zero-Knowledge Scalable Transparent ARguments of Knowledge (zk-STARKs) (EIP-4844: Shard Blob Transactions, n.d.).

One repeated concern surrounding cryptocurrency is the effect that it has on the environment. Not only this, but as society integrates towards a fuller integration between the physical and digital worlds, there will be many interoperable blockchains. Consider just the instance of a soulbound NFT tied to a driver's license, then multiply that by the number of people who have their license. There is an electrical cost associated with each minting of a new NFT as well as with each presentation, verification, etc. When these actions are recorded onto the blockchain, there is the expense associated with including it in the block and mining the entire block. Processes such as Ethereum's PoS mechanism help to reduce this burden. It's projected that the electricity demands for NFTs are expected to increase in the near future but will then start to decrease until its energy consumption in 2050 is lower than it was back in 2021 (Zhao & You, 2023). In fact, it's projected that the entire metaverse sector can reduce the domestic energy usage in the United States by 92 EJ before 2050 and can reduce air pollutants by 10-23% (Zhao & You, 2023).

In a fully distributed financial market, trading occurs peer-to-peer which opens the door for participants to trade around the clock. However, this is not a perfect system. For example, look at mutual funds. A mutual fund is essentially a basket of securities which are actively managed by one or more individuals in return for a fee. Currently in the United States, they are priced once a day at market close, and buy/sell orders are executed based on the calculation of the Net Asset Value (NAV) per share. Opposed to selling stock to another individual, when a mutual fund is sold, it is redeemed by the fund and paid out in cash. If trading happens at all times of the week, this adds pressure to mutual fund companies (or any institution involved in the financial markets in some way) to manage their portfolio consistently.

In addition to the above, there remain legal, cultural, and religious obstacles for the full adoption of blockchain, NFTs, and the metaverse. On the legal side, the nature of its decentralized architecture makes it rather difficult for centralized governments/entities to exert control on it as well as the applications built on top of it. However, a legal framework is slowly starting to develop, as is evidenced through the sale of an NFT not conveying copyright ownership (Bolton & Cora, 2021). On the cultural side, there are groups, such as the First Nations and other indigenous people, who make the case that virtual land is being transacted with disregard to the cultural significance of being a landowner (Barba et al., 2022). Looking at the religious challenge to widespread blockchain, NFT, and metaverse adoption, Islamic finance is a case study. For example, an Ijara is a "Shariah-compliant version of a conventional lease" for physical/digital property that lays out aspects such as designating who is responsible for maintenance obligations, insurance of the property, and the responsibility in the event of the loss or destruction of the property (Katterbauer et al., 2022). To be fully embraced by society, not only must this technology overcome the technical hurdles, but also the intangible ones.

## 6. Conclusion

As technology continues to progress, today marks a pivotal point between traditional and futuristic living. Though digital reality has been around for an extended period through various video games and online chat rooms, it is only in more recent times that the digital and physical worlds start to mesh through the introduction of virtual and augmented reality. When DLT is added to the mix, a potent mixture that is ripe for innovation and explosion is created. This creates an atmosphere of uncertainty and excitement as society holds its collective breath to see if, where, and how this innovative technology will be implemented in their daily lives. Initially supported by a small subsection of the population, NFTs grew to an enormous albeit speculative bubble, but moving past this initial hype, they started to find their real utilitarian value as they can be implemented in all sorts of unique and innovative ways.

A major concern regarding the implementation of a blockchain at a governmental level is the technical limitations of the blockchain. To foster the collective needs of a society, the blockchain which is to be the backbone must support a very high level of transactions with a minimal cost structure and large amounts of data. While mechanisms such as Ethereum's PoS system and the IPFS system help to alleviate some of these problems, they also introduce other issues to consider. Additionally, there must be interoperability amongst the different blockchains so that users within any country can substitute their local central bank digital currency with that of another type or token.

While the full implementation of blockchain, NFTs, and the metaverse within digital government and taking place on a global scale is sure to be a long, arduous process, it is sure to be well worth it in the end should it be implemented with a great amount of forethought. Doing so will increase transparency, data management and retrieval, distributed collaboration, reduce interaction friction, and reduce energy consumption while providing for a more interconnected populace and a higher productivity level with a reduction in the burden taken on by the average citizen. It will also help to reduce the amount and frequency of fraud as all transactions will be posted onto the blockchain. However, this last point will likely be one of contention as some groups of people, who are untrusting of the government, will likely push back on this adoption. To their point of view, cryptocurrency was originally created as an alternative to the fiat systems already in place, and it was to offer a level of anonymity in transactions not afforded by current systems.

As the idea of a smart city starts to take hold, society will be able to realize the full benefit of DLT, NFTs, Artificial Intelligence, and the Internet of Things (IoT), as well as some issues that are still unknown or realized. Some societal issues still need answering, such as if and to what extent will smart cities close/widen the gap between those in poverty and the affluent. Likewise, how does this same concept extend on a national and global scale between the wealth of cities within a country and between countries respectively. Could this result in a migration of wealth into or out from a specific geographic location? Could this then subsequently rebalance the control of power on the global stage? For smart cities to be effective in the long run, these are all questions that need to be answered prior to their complete implementation. In conclusion, blockchain technology and NFTs are just as exciting as they are complicated. They offer a brand-new type of solution to many old problems, both within and outside of digital government. While this technology provides a great opportunity, it is wise to approach it carefully. By erring on the side of caution and in a conservative manner, this will enable the safe and sound development and adoption of DLT by the masses. Only time will tell exactly how and to what extent they will be implemented, as the possibilities are great, so are the potential rewards.